\documentclass[12pt]{article}

plus3pt minus2pt \abovedisplayshortskip=12pt plus3pt minus2pt
\usepackage{graphics, amsfonts, amssymb, latexsym, amsmath, amsxtra}
\usepackage[dvips]{graphicx}

\newtheorem{theorem}{Theorem}

\newcommand {\mc} {\mathcal}

\begin{document}

\begin{center}

\textbf{\Large{ Volterra's realization of the KM-system}}

\vskip 1cm

\noindent Maria A. Agrotis, Pantelis A. Damianou

\vskip .4cm

\noindent \footnotesize{\textit{Department of Mathematics and
Statistics, University of Cyprus, Nicosia 1678, Cyprus}}

\vskip 1cm

\noindent \textbf{Abstract}

\end{center}

{\footnotesize{We construct a symplectic realization of the
KM-system and obtain the higher order Poisson tensors and
commuting flows via the use of a recursion operator. This is
achieved by doubling the number of variables through Volterra's
coordinate transformation. An application of Oevel's theorem
yields master symmetries, invariants  and deformation relations.}}

\vskip .3cm

\noindent MSC 37J35, 70H06

\vskip .3cm

\noindent keywords: Volterra system, recursion operator,
symplectic structure.

\vskip .3cm

\noindent Corresponding author's email: agrotis@ucy.ac.cy

\vskip .3cm

\section{Introduction} \label{intro}

The KM-system, also known as Volterra system, is defined by

\begin{equation}
\dot u_i = u_i(u_{i+1}-u_{i-1}) \qquad i=1,2, \dots,n, \label{a1}
\end{equation}
where $u_0 \!= u_{n+1} \!=0$. It has been used as a model for
predator-prey evolution systems \cite{volterra}, as well as a
discretization of the Korteweg-de Vries equation. The
integrability of the system was established in \cite{kac} and
\cite{moser2}. In \cite{kac} the inverse scattering technique is
formulated in a discrete setting and applied on equations
\eqref{a1} to produce explicit solutions. Moser using a different
method, namely continued fractions, has also integrated the model.

A diffeomorphism is established between the KM-system (\ref{a1})
and the celebrated Toda lattice equations
\begin{eqnarray*}
&& \dot{a}_i=a_i (b_{i+1}-b_{i}) \hspace{.9cm} i=1,\ldots,n-1 \\
&& \dot{b}_i=2(a_i^2-a_{i-1}^2)  \hspace{.95cm} i=1,\ldots,n,
\end{eqnarray*}
via  a transformation of H\'{e}non
\begin{eqnarray*}
&& a_i=-\frac{1}{2}\sqrt{u_{2i}u_{2i-1}}  \hspace{1.4cm}i=1,\ldots,n-1 \\
&& b_i=\frac{1}{2}(u_{2i-1}+u_{2i-2}) \hspace{.9cm} i=1,\ldots,n.
\end{eqnarray*} Thus, the hierarchy
of Poisson tensors, Hamiltonian functions, constants of motion and
master symmetries known for the Toda lattice and expressed in
Flaschka's coordinates $(a,b)$, can be mapped to the corresponding
ones for the KM-system in $u-$coordinates \cite{damianou1}. We
note that the number of variables for the Toda lattice is odd and
therefore we restrict our attention to the Volterra system with an
odd number of variables.

A Hamiltonian description of the Volterra model can be found in
the book of Fadeev and Takhtajan  \cite{fadeev}. Later on, in
\cite{damianou1} two polynomial Poisson tensors of degree two and
three are considered and placed in an infinite sequence of Poisson
tensors that satisfy Lenard type relations. The quadratic Poisson
bracket, $\pi_2$, is defined by the formulas
\begin{equation}
\{u_i,u_{i+1}\}=u_i u_{i+1},
\end{equation} and all other brackets are zero. Using
$H=\sum_{i=1}^{2n-1} u_i$ as the Hamiltonian and the Poisson
bracket  $\pi_2$, the Volterra equations are written in the form,
$\dot{u_i}=\{u_i,H\}.$

We will follow \cite{damianou1} and use the Lax pair of that
reference. It has the advantage of making the equations
homogeneous polynomial. The Lax pair is given by
\begin{equation}
\dot{L}=[B, L],
\end{equation}
where

\begin{equation}
L= \begin{pmatrix} u_1 & 0 & \sqrt{u_1 u_2} & 0 & \dots & \  & 0
\cr 0 &
u_1 +u_2 & 0& \sqrt{u_2 u_3}  & & & \vdots \\
 \sqrt{u_1 u_2} & 0 & u_2 +u_3 & &  \ddots & &  \\
 0 & \sqrt{u_2 u_3} & &  & & & \\
  \vdots & & \dots & & & & \sqrt{u_{2n-2} u_{2n-1}} \\
 & & & & & u_{2n-2}+u_{2n-1} & 0 \\
 & & & & \sqrt{u_{2n-2} u_{2n-1}} &0& u_{2n-1} \end{pmatrix}
 \nonumber
\end{equation}
and

\begin{equation}
B=\begin{pmatrix} 0 & 0 & { 1 \over 2} \sqrt{u_1 u_2} & 0 & \dots
&
\ & 0 \cr 0 & 0 & 0&{ 1 \over 2} \sqrt{u_2 u_3}  & & & \vdots \\
-{ 1\over 2} \sqrt{u_1 u_2} & 0 & 0 & &  \ddots & &  \\ 0 & -{
1\over 2} \sqrt{u_2 u_3} & &  & & & \\ \vdots & & \dots & & & & {1
\over 2} \sqrt{u_{2n-2} u_{2n-1}} \cr & & & & & 0 & 0 \\ & & & &-{
1 \over 2}  \sqrt{u_{2n-2} u_{2n-1}} &0& 0
\end{pmatrix}. \nonumber
\end{equation}

\noindent This is an example of an isospectral deformation; the
entries of $L$ vary over time but the eigenvalues  remain
constant. It follows that the  functions $ H_i={1 \over i} {\rm
Tr} \, L^i$ are constants of motion.

The cubic Poisson bracket, which corresponds to the second KdV
bracket in the continuum limit, is defined by

\begin{equation} \nonumber
\begin{array}{rcl}
\{ u_i, u_{i+1} \}&=& u_i u_{i+1} (u_i+ u_{i+1}) \\
\{ u_i, u_{i+2} \} &=& u_i u_{i+1} u_{i+2} \, ,
\end{array}
\end{equation}
and all other brackets are zero. We denote this bracket by
$\pi_3$. The Lenard relations take the form

\begin{equation*}
 \pi_3\, \nabla
H_i = \pi_2 \, \nabla H_{i+1}.
\end{equation*} The higher order Poisson
brackets are constructed using a sequence of master symmetries
$Y_i, \: \: i=0,1, \dots$. We define $Y_0$ to be the Euler vector
field

\begin{equation} \nonumber
Y_0=\sum_{i=1}^{2n-1} u_i {\partial \over \partial u_i} ,
\end{equation}
and $Y_1$ the master symmetry
\begin{equation} \nonumber
Y_1=\sum_{i=1}^{2n-1} U_i {\partial \over \partial u_i} \ ,
\end{equation}
where
\begin{equation}\nonumber
U_i=(i+1) u_i u_{i+1} +u_i^2+(2-i) u_{i-1} u_i \ .
\end{equation}
One can verify that the bracket $\pi_3$ is obtained from $\pi_2$
by taking the Lie derivative in the direction of $Y_1$.

The brackets $\pi_2$ and $\pi_3$ are just the beginning of an
infinite family constructed in \cite{damianou1} using master
symmetries. We quote the result:

\begin{theorem} \label{th1}
There exists a sequence of Poisson tensors $\pi_j$ and a sequence
of master symmetries $Y_j$ such that:

\smallskip

\noindent {\it i)} $\pi_j$ are all Poisson.

\smallskip

\noindent {\it ii)} The functions $H_i$  are in involution
 with respect to all of the $\pi_j$.

\smallskip

 \noindent
 {\it iii)} $Y_i (H_j) =(i+j) H_{i+j} $.

\smallskip

\noindent {\it iv)} $L_{Y_i} \pi_j =(j-i-2) \pi_{i+j} $.

\smallskip

\noindent {\it v)} $[Y_i, \ Y_j]=(j-i)Y_{i+j}$.

\smallskip

\noindent {\it vi)} $\pi_j \nabla H_i =\pi_{j-1} \nabla H_{i+1} $,
where $\pi_j$ denotes  the Poisson matrix  of the tensor $\pi_j$.
\end{theorem}

The KM-system  is a special case of the more general
Lotka-Volterra system which  has the form,

\begin{equation}
\dot u_i = \sum_{k=1}^{N} a_{ik} u_i u_{k} \qquad i=1,2, \ldots, N
, \label{a2}
\end{equation}
where $(a_{ij})$ is a fixed matrix.

 In the early
work on (\ref{a2}), Volterra introduced a transformation from
$\mathbb{R}^{2N}$ to $\mathbb{R}^{N},$ in his attempt to provide a
Hamiltonian formulation; see for example \cite{fernandes}.
Specifically he doubled the number of variables by defining

\begin{eqnarray}
q_i(t)&=&\int_{0}^{t} u_i(\tau) d\tau \\
p_i(t)&=& \ln(\dot{q_i}) -\frac{1}{2} \sum_{k=1}^{N} a_{ik} q_k,
\end{eqnarray} i=1,\ldots, N, for a skew-symmetric $(a_{ij}).$

The explicit form of Volterra's transformations from
$\mathbb{R}^{2N}$ to $\mathbb{R}^{N},$  is

\begin{equation} u_i=e^{p_i+\frac{1}{2}\sum_{k=1}^{N} a_{ik} q_k}  \ \ \ \
i=1,2, \dots, N \ . \end{equation} The Hamiltonian function is
given by

\begin{equation}
H=\sum_{i=1}^{N}\dot{q}_i=\sum_{i=1}^{N}u_i ,
\end{equation} which takes the form
\begin{equation}
H=\sum_{i=1}^{N}e^{p_i+\frac{1}{2}\sum_{k=1}^{N} a_{ik} q_k}.
\end{equation}  System (\ref{a2}) can then be endowed with the
following symplectic realization

\begin{eqnarray}
&& \dot{q}_i=\frac{\partial H}{\partial p_i}=\{q_i,H \} \\
&& \dot{p}_i=-\frac{\partial H}{\partial q_i}=\{p_i,H\},
\end{eqnarray} where the Poisson bracket in $(q,p)$ coordinates in
$\mathbb{R}^{2N}$ is the canonical one. We note that for the
KM-system in $u$-space both Poisson tensors $\pi_2$ and $\pi_3$
are degenerate. Therefore, an application of the theory of
recursion operators  is hindered.

In this paper we consider the Volterra model in
$\mathbb{R}^{2n-1}$ and obtain a symplectic realization of the
system by increasing the dimension of the space. Namely, the
number of variables is doubled through Volterra's coordinate
transformation. We rediscover the higher order Poisson tensors and
flows for the system via the use of a recursion operator. We
define a conformal symmetry in symplectic space and apply Oevel's
theorem to produce deformation relations, which can then be
projected to give the deformation relations for the Volterra
system in $\mathbb{R}^{2n-1}.$

\section{Master Symmetries and Recursion Operators}
\label{symmetries}

Let us consider the differential equation $\dot{x}=\mathcal{X}(x)$
on a manifold $M$ defined by the Hamiltonian vector field
$\mathcal{X}$. Below we give the definition of  master symmetries,
due to Fokas and Fuchssteiner \cite{fokas1}, and briefly mention
their basic properties. A vector field $Z$ is a symmetry of the
equation if $[Z,\mathcal{X}]=0.$ In the case that $Z=Z(t,x)$, $Z$
is a time-dependent symmetry if

\begin{equation*}
\frac{\partial Z}{\partial t}+[Z,\mathcal{X}]=0.
\end{equation*}
A more general definition is that of a generator of symmetries.
$Z$ is called a generator of degree zero if $[Z,\mathcal{X}]=0,$
and a generator of degree one if
$[[Z,\mathcal{X}],\mathcal{X}]=0$. A generator of degree $k$ is
the one that satisfies $[[\ldots[Z,\mathcal{X}],\ldots]=0,$ where
there are $k+1$ nested Lie brackets. We remark that if $Z$ is a
generator of degree $k$ then $[Z,\mc{X}]$ is a generator of degree
$k-1.$ Also, if $Z$ is a generator of degree $k$ then $Z$ is a
generator of degree $i\ge k.$ A symmetry is a generator of degree
zero. A generator of degree one that is not a generator of degree
zero is called a master symmetry. Oevel's theorem provides a
useful method  for constructing master symmetries.

Suppose that we have a bi-Hamiltonian system \cite{magri} with a
symplectic Poisson tensor. Namely, a pair of Poisson tensors $J_0$
and $J_1$, with $J_0$ symplectic and a pair of Hamiltonian
functions $H_1,H_2$ that give rise to the same system, i.e,

\begin{equation}
J_0 \nabla H_2=J_1 \nabla H_1.
\end{equation} Then a recursion operator $\mc{R}$ is defined
by $\mc{R}=J_1 J_0^{-1},$ and gives rise to a family of
Hamiltonian vector fields that are defined recursively as,

\begin{equation*}
\mc{X}_{i} = {\cal R}^{i-1} \mc{X}_1 ,
\end{equation*}
and higher order Poisson tensors \begin{equation} J_i = {\mc
R}^{i} J_0 . \end{equation} The Hamiltonians $H_i$ corresponding
to the vector fields $\mc{X}_i$ are given by $\nabla
H_i=(\mc{R}^{*})^i \nabla H_0$. These higher order flows have a
multi-Hamiltonian formulation
\begin{equation}
\mc{X}_{i+j}=J_i \nabla H_j.
\end{equation}
Magri's theorem \cite{magri} states that the flows $\mc{X}_i$
pairwise commute. Also the functions $H_i$ are constants of motion
for each flow and commute with respect to all higher order Poisson
tensors. We thus have an infinite sequence of involutive
Hamiltonian flows. Furthermore, Oevel's theorem provides a method
for constructing master symmetries \cite{oevel2}. We quote the
theorem.

\begin{theorem} \label{th2}
Suppose that   $X_0$ is a conformal symmetry for both  $\pi_1$,
$\pi_2$ and $H_1$, i.e.  for some scalars $\lambda$, $\mu$, and
$\nu$ we have \begin{equation} {\cal L}_{X_0} \pi_1= \lambda
\pi_1, \quad{\cal L}_{X_0} \pi_2 = \mu \pi_2, \quad {\cal L}_{X_0}
H_1 = \nu H_1  . \end{equation} Then the vector fields $X_i =
{\cal R}^i X_0$ are master symmetries and we have,
\begin{eqnarray*}
&& (a) \ {\cal L}_{X_i} H_j = (\nu +(j-1+i) (\mu -\lambda))
H_{i+j}
\\
&& (b) \ {\cal L}_{X_i} \pi_j = (\mu +(j-i-2) (\mu -\lambda))
\pi_{i+j} \\
&& (c) \ [X_i, X_j]= (\mu - \lambda) (j-i) X_{i+j}  \ .
\end{eqnarray*}
\end{theorem}
As a corollary to Oevel's theorem we have the existence of the
following time-dependent symmetries for each flow in the
hierarchy,

\begin{equation}
Y_{\mc{X}_i}=X_i + t(\mu+\nu+(j-1)(\mu-\lambda))
\mc{X}_{i+j},\quad i,j=1,2,\ldots
\end{equation}
In the next section we will formulate the bi-Hamiltonian Volterra
system in a symplectic setting so that we can apply the theory
described in this section and obtain the results stemming out of
the theorems of Magri and Oevel.

\section{Symplectic setting} \label{symplectic}

We consider the Volterra map

\begin{eqnarray}
&& \Psi: {\bf R}^{2(2n-1)} \mapsto {\bf R}^{2n-1} \nonumber \\
&& u_{i} = e^{p_i+\frac{1}{2}(q_{i+1}- q_{i-1})}  \qquad
i=1,\ldots,2n-1 \label{u-variables} ,
\end{eqnarray} where $q_0=q_{2n}=0.$ We note that $u_0=u_{2n}=0.$
The Hamiltonian in $(q,p)$ coordinates is given by

\begin{equation} h_1= \sum_{i=1}^{2n-1} e^{p_i+\frac{1}{2}(q_{i+1}-q_{i-1})},
\label{b2} \end{equation} and together with the canonical
symplectic bracket in ${\bf R}^{2(2n-1)}$, call it $J_2$,
corresponds  to the Volterra system (\ref{a1}) under the mapping
(\ref{u-variables}). In particular, the degenerate quadratic
Poisson tensor $\pi_2$ defined in Section \ref{intro}, is lifted
to the symplectic bracket $J_2$ via transformation
(\ref{u-variables}). To find the pre-image of the cubic bracket
$\pi_3$, we will lift the master symmetry $Y_1$ of Section
\ref{intro} from the u-space in ${\bf R}^{2n-1}$ to a master
symmetry $X_1$ in  the symplectic space $(q,p) \in {\bf
R}^{2(2n-1)}.$ In fact, $\mc{L}_{X_1}J_2=J_3,$ where $X_1$
projects to $Y_1$ using the Volterra map. One possible definition
for $X_1$ is the following:

\begin{equation} X_1 = \sum_{i=1}^{2n-1} A_i {\partial \over \partial q_i}  +
\sum_{i=1}^{2n-1} B_i{ \partial \over \partial p_i} \,
\end{equation} where,
\begin{eqnarray*}
A_i &=& \sum_{j=1}^{2n-1} c_{j,i} \: e^{p_j+\frac{1}{2}(q_{j+1}-
q_{j-1})},\\
B_i &=& (i+1) \: e^{p_{i+1}+\frac{1}{2}(q_{i+2}- q_{i})} +
e^{p_i+\frac{1}{2}(q_{i+1}- q_{i-1})} + (2-i) \:
e^{p_{i-1}+\frac{1}{2}(q_{i}- q_{i-2})}  \\
& & + \frac{1}{2} \sum_{j=1}^{2n-1} (c_{j,i-1}-c_{j,i+1}) \:
e^{p_j+\frac{1}{2}(q_{j+1}- q_{j-1})}, \label{b1}
\end{eqnarray*} for $i=1,2, \dots, 2n-1.$ The constants $c_{i,j}$ are given
by
\begin{eqnarray*}
&& c_{i,j}=0, \hspace{1.45cm} i=1,\ldots,2n-2, \:\: j>i \\
&& c_{i,j}=-1, \hspace{1.1cm} i=2,\ldots,2n-1, \:\: j<i \\
&& c_{i,i}=i-1, \hspace{.8cm} i=1,\ldots,2n-1.
\end{eqnarray*} We note that $c_{j,0}=c_{j,2n}=0.$ The constant matrix
$C:=(c_{i,j})$ takes the form,

\begin{equation*}
C:=\begin{pmatrix} 0 & 0 & 0 & 0 & \ldots & 0 \cr
                  -1 & 1 & 0 & 0 & \ldots & 0 \cr
                  -1 & -1 & 2 & 0 & \ldots & 0 \cr
                  \vdots & \vdots & \ddots & \ddots & \ddots & \vdots \cr
                  -1 & \ldots & \ldots & -1 & 2n-3 & 0 \cr
                  -1 & \ldots & \ldots & \ldots & -1 & 2n-2 \end{pmatrix} .
\end{equation*}

Taking the Lie derivative of the symplectic bracket $J_2$ in the
direction of $X_1$ we obtain the Poisson bracket $J_3$,
\begin{eqnarray}
&& \{ q_i, q_{j} \}  = e^{p_j+\frac{1}{2}(q_{j+1}- q_{j-1})}
\hspace{5.2cm} i
< j \label{J3bracket} \\
&& \{ q_{1}, p_1 \} =  e^{p_1+\frac{1}{2}q_{2}}+\frac{1}{2}
e^{p_2+\frac{1}{2}(q_{3}- q_{1})} \nonumber \\
&& \{ q_{i}, p_i \} = \frac{1}{2}
e^{p_i+\frac{1}{2}(q_{i+1}-q_{i-1})}+\frac{1}{2}
e^{p_{i+1}+\frac{1}{2}(q_{i+2}- q_{i})}  \hspace{1.7cm} i=2,\ldots,2n-1 \nonumber \\
&& \{ q_{i}, p_{i+1} \} = \frac{1}{2}
e^{p_{i+2}+\frac{1}{2}(q_{i+3}- q_{i+1})}  \hspace{4.3cm} i=1,\ldots,2n-2 \nonumber \\
&& \{ q_{2}, p_{1} \} =
e^{p_{2}+\frac{1}{2}(q_{3}- q_{1})}  \nonumber \\
&& \{ q_{i}, p_{i-1} \} = \frac{1}{2}
e^{p_{i}+\frac{1}{2}(q_{i+1}- q_{i-1})}  \hspace{4.7cm} i=3,\ldots,2n-1 \nonumber \\
&& \{ q_{i}, p_{j} \} = -\frac{1}{2} e^{p_{j-1}+\frac{1}{2}(q_{j}-
q_{j-2})} + \frac{1}{2}
e^{p_{j+1}+\frac{1}{2}(q_{j+2}- q_{j})}  \hspace{1.2cm} j \ge i+2 \nonumber \\
&& \{ q_{i}, p_{1} \} = \frac{1}{2}
e^{p_{i}+\frac{1}{2}(q_{i+1}- q_{i-1})}  \hspace{5cm} i=3,\ldots,2n-1 \nonumber \\
&& \{ p_{1}, p_{2} \} = \frac{1}{2}
e^{p_{1}+\frac{1}{2}q_{2}}+\frac{1}{4} e^{p_{2}+\frac{1}{2}(q_{3}-
q_{1})}-\frac{1}{4} e^{p_{3}+\frac{1}{2}(q_{4}- q_{2})}    \nonumber \\
&& \{ p_i , p_{i+1} \} = \frac{1}{4}
 e^{p_{i}+\frac{1}{2}(q_{i+1}- q_{i-1})}+\frac{1}{4}
 e^{p_{i+1}+\frac{1}{2}(q_{i+2}- q_{i})}  \hspace{1.4cm} i=2, \dots, 2n-2
 \nonumber \\
&& \{ p_{1}, p_{3} \} = \frac{1}{2}
 e^{p_{2}+\frac{1}{2}(q_{3}- q_{1})}-\frac{1}{4}
 e^{p_{4}+\frac{1}{2}(q_{5}- q_{3})}   \nonumber \\
&& \{ p_i , p_{i+2} \} = \frac{1}{4}
 e^{p_{i+1}+\frac{1}{2}(q_{i+2}- q_{i})}  \hspace{4.7cm} i=2, \dots, 2n-3
 \nonumber \\
&& \{ p_1 , p_{j} \} = -\frac{1}{4}
e^{p_{j+1}+\frac{1}{2}(q_{j+2}- q_{j})} +  \frac{1}{4}
e^{p_{j-1}+\frac{1}{2}(q_{j}- q_{j-2})} \hspace{1.2cm} j = 4,
\ldots, 2n-1 \nonumber,
\end{eqnarray} and all other brackets are zero. We recall that
$e^{p_{2n}+\frac{1}{2}(q_{2n+1}-q_{2n-1})}=u_{2n}=0.$ The Jacobi
identity for the bracket $J_3$ can be rigorously checked by
considering the following four cases: (a) three $q$, (b) three
$p$, (c) two $p$ and one $q$, and (d) two $q$ and one $p$. For
example, the Jacobi identity for $\: q_i,q_j,q_k\:$ for $\: 1 \le
i <j<k \le 2n-1 \:$ can be broken up to two subcases: $a1)\:
k=j+1,$ and $a2) \: k \ge j+2.$ In a similar manner one can
consider the other three cases.

Under the Volterra transformation, $J_2$ maps to $\pi_2$ and $J_3$
to $\pi_3$. The function
\begin{equation*}
h_2=\frac{1}{2} \sum_{i=1}^{2n-1} e^{2p_i+q_{i+1}-q_{i-1}}+
\sum_{i=1}^{2n-2} e^{p_{i}+p_{i+1}+\frac{1}{2}
(q_{i+2}+q_{i+1}-q_{i}-q_{i-1})}
\end{equation*} corresponds under mapping (\ref{u-variables}) to a
constant multiple of $H_2=\frac{1}{2} \mbox{Tr}(L)^2$. We recall
that $H_1,H_2$ and $\pi_2,\pi_3$ constitute a bi-Hamiltonian pair,
\begin{equation}
\pi_2 \nabla H_2= \pi_3 \nabla H_1.
\end{equation}
However, both Poisson tensors are degenerate.
The Volterra map places this bi-Hamiltonian pair in a symplectic
setting. That is,
\begin{equation}
J_2 \nabla h_2= J_3 \nabla h_1.
\end{equation}
Therefore, the definition of a recursion operator, $\mc{R}=J_3
J_2^{-1}$ is possible. $ J_3$ is by construction compatible with
$J_2$ since it is generated from a master symmetry; see
\cite{damianou2}. We note the absence of a negative recursion
operator as in \cite{marmo}  using this method, since  the matrix
representing $J_3$ is not invertible.

A multi-Hamiltonian structure of the form, $\mc{X}_{i+j}=J_i
\nabla h_j,$ is provided by the higher order Poisson tensors  and
Hamiltonian vector fields
\begin{equation}
J_i=\mc{R}^{i-2}J_2, \quad i=3,4,\ldots,
\end{equation}
\begin{equation}
\mc{X}_i=\mc{R}^{i-1}\mc{X}_1, \quad i=2,3,\ldots,
\end{equation}
where $\mc{X}_i$ stands for $\mc{X}_{h_i}.$

Theorem \ref{th2} requires the existence of a conformal symmetry
$X_0$ such that

\begin{equation} {\cal L}_{X_0} J_2=\lambda J_2, \quad
{\cal L}_{X_0} J_3= \mu J_3, \quad {\cal L}_{X_0}(h_1)=\nu h_1.
\label{cs}
\end{equation} We define the conformal symmetry

\begin{equation}
X_0=\sum_{i=1}^{2n-1} \frac{\partial}{\partial p_i},
\end{equation} and one can check that relations (\ref{cs}) are satisfied with
$\lambda=0,\:\,\mu=1,\:\,\nu=1.$ Therefore, in addition to the
infinite family of commuting Hamiltonian flows, we have the
following deformation relations:

\begin{equation}
[X_i, h_j]= (i+j)h_{i+j}
\end{equation}

\begin{equation}
\hspace{.35cm} L_{X_i}  J_j = (j-i-2) J_{i+j}
\end{equation}

\begin{equation}
\hspace{.2cm}[ X_i, X_j ]  = (j-i) X_{i+j} .
\end{equation} Using the
Volterra map we can project these to the $u-$space and provide an
alternative proof of the statements of Theorem \ref{th1}.

\section{Discussion}

A different symplectic realization for the KM-system has been
achieved recently  in \cite{marmo} using the map
\begin{eqnarray}\label{trans1}
&& \Phi: \mathbb{R}^{2n} \longmapsto \mathbb{R}^{2n-1}\\
&& u_{2i-1} = - e^{p_i} \qquad i=1,\ldots,n  \nonumber\\
&& u_{2i} =  e^{q_{i+1}- q_{i}}  \qquad i=1,\ldots,n-1. \nonumber
\end{eqnarray} The Hamiltonian is defined as

\begin{equation}
H=-\sum_{i=1}^{n}e^{p_i}+\sum_{i=1}^{n-1}e^{q_{i+1}-q_{i}}
\end{equation} and the standard symplectic bracket in
$(q,p)-$space maps to the degenerate quadratic Poisson tensor
$\pi_2$ via transformation (\ref{trans1}). A second symplectic
bracket is obtained by lifting the cubic bracket $\pi_3.$

In this paper we consider the Volterra map
\begin{eqnarray}
&& \Psi: {\bf R}^{2(2n-1)} \mapsto {\bf R}^{2n-1} \label{trans2} \\
&& u_{i} = e^{p_i+\frac{1}{2}(q_{i+1}- q_{i-1})}  \qquad
i=1,\ldots,2n-1, \nonumber \end{eqnarray} in order to lift the
bi-Hamiltonian structure of the KM-system to a symplectic space in
$\mathbb{R}^{2(2n-1)}.$ The big difference between the dimensions
of the source and the target space in \eqref{trans2} impedes the
application of the methodology used in \cite{marmo}. However, the
existence of a pair of Poisson tensors, at least one of which is
non-degenerate, is possible. A second bracket $J_3$ in
$\mathbb{R}^{2(2n-1)}$ is obtained so that its image under mapping
\eqref{trans2} is the cubic bracket $\pi_3.$ Since $J_2$ is
symplectic, a recursion operator is defined as $\mc{R}=J_3
J_2^{-1},$  and used to give rise to an infinite hierarchy of
commuting Hamiltonian flows and Poisson tensors. The conformal
symmetry of the KM-system in $u-$space is lifted to the symplectic
$(q,p)-$space, and an application of Oevel's theorem leads to an
infinite number of master symmetries, Poisson tensors and
invariants. We note the absence of a negative recursion operator
using this realization since  $J_3$ is non-invertible.

\section{Acknowledgements}
One of the authors M.A.A would like to thank the Cyprus Research
Promotion Foundation for support through the grant CRPF0504/03.

\end{document}